\documentclass[onecolumn,showpacs,superscriptaddress]{revtex4}
\usepackage{graphicx}
\usepackage{amsmath}
\usepackage{amsthm}
\usepackage{amssymb}
\usepackage{latexsym}
\usepackage{array}
\usepackage{float}
\usepackage{amsfonts}
\usepackage{mathrsfs}
\usepackage{verbatim}

\begin{document}

\title{A quantum algorithm for obtaining the lowest eigenstate of a Hamiltonian assisted with an ancillary qubit system}

\author{Jeongho Bang}\email{jbang@snu.ac.kr}
\affiliation{Center for Macroscopic Quantum Control, Department of Physics and Astronomy, Seoul National University, Seoul, 151-747, Korea}
\affiliation{Department of Physics, Hanyang University, Seoul 133-791, Korea}

\author{Seung-Woo Lee}
\affiliation{Center for Macroscopic Quantum Control, Department of Physics and Astronomy, Seoul National University, Seoul, 151-747, Korea}

\author{Chang-Woo Lee}
\affiliation{Department of Physics, Texas A\&M University at Qatar, PO Box 23874, Doha, Qatar}
\affiliation{Center for Macroscopic Quantum Control, Department of Physics and Astronomy, Seoul National University, Seoul, 151-747, Korea}

\author{Hyunseok Jeong}
\affiliation{Center for Macroscopic Quantum Control, Department of Physics and Astronomy, Seoul National University, Seoul, 151-747, Korea}
\affiliation{Centre for Quantum Computation and Communication Technology, School of Mathematics and Physics, University of Queensland, Brisbane, Queensland 4072, Australia}

\received{\today}

\begin{abstract}
We propose a quantum algorithm to obtain the lowest eigenstate of any Hamiltonian simulated by a quantum computer. The proposed algorithm begins with an arbitrary initial state of the simulated system. A finite series of transforms is iteratively applied to the initial state assisted with an ancillary qubit. The fraction of the lowest eigenstate in the initial state is then amplified up to $\simeq 1$. We prove that our algorithm can faithfully work for any arbitrary Hamiltonian in the theoretical analysis. Numerical analyses are also carried out. We firstly provide a numerical proof-of-principle demonstration with a simple Hamiltonian in order to compare our scheme with the so-called ``Demon-like algorithmic cooling (DLAC)'', recently proposed in [Nature Photonics {\bf 8}, 113 (2014)]. The result shows a good agreement with our theoretical analysis, exhibiting the comparable behavior to the best `cooling' with the DLAC method. We then consider a random Hamiltonian model for further analysis of our algorithm. By numerical simulations, we show that the total number $n_c$ of iterations is proportional to $\simeq {\cal O}(D^{-1}\epsilon^{-0.19})$, where $D$ is the difference between the two lowest eigenvalues, and $\epsilon$ is an error defined as the probability that the finally obtained system state is in an unexpected (i.e. not the lowest) eigenstate.
\end{abstract}


\maketitle

\newcommand{\bra}[1]{\left<#1\right|}
\newcommand{\ket}[1]{\left|#1\right>}
\newcommand{\abs}[1]{\left|#1\right|}
\newcommand{\expt}[1]{\left<#1\right>}
\newcommand{\braket}[2]{\left<{#1}|{#2}\right>}
\newcommand{\ketbra}[2]{\left|{#1}\right>\left<{#2}\right|}
\newcommand{\commt}[2]{\left[{#1},{#2}\right]}

\newcommand{\identity}{1\!\!1}

%----------------------------
\section{Introduction}\label{sec:1}

It is essential to deal with complex systems whose dynamics are typically described by many-body Hamiltonians in various fields of quantum information and computation. One important issue that has been raised in quantum computation is the eigen-problem of a Hamiltonian whose Hilbert-space is very large (possibly, even several hundreds or thousands). In particular, one of the frequently faced but quite formidable problem \footnote{Such a task belongs to the class of ``Non-deterministic Polynomial'' (NP), or its quantum generalization, called ``Quantum-Merlin-Arthur'' (QMA) \cite{Liu07}.} is how to find the lowest energy eigenstate of a complex system \cite{Fischer91,Rojas96,Albert02}.

A possible way to attack this problem is to use an approach of so-called algorithmic quantum cooling (AQC), which is often referred to as a systematic technique of descreasing (increasing) the fractions of the higher (lower) energy eigenstates of the system \cite{Boykin02,Fernandez04,Verstraete09,Xu14}. The AQC is particularly useful when it is desired to initialize the part of higher energy states, say `hot' qubits (e.g., a macroscopic number of spins \cite{Gershenfeld97}), to lower energy states in ensemble quantum computation \cite{King98,Roos99,James00,Popp06,Christ07} or quantum simulation \cite{Lewenstein07,Bloch12}. Most recently, a simple but powerful `pseudo' AQC method, ``Demon-like algorithmic cooling (DLAC)'', has been proposed and experimentally demonstrated \cite{Xu14}. The DLAC method can drive a given initial state to the lowest eigenstate using a quantum-circuit module with an ancillary qubit. The core process in this method is the measurement of the ancillary qubit to discard the `heated' state and to leave only the `cooled' state (like Maxwell's famous `demon' \cite{Leff90,Lloyd97}).

In this work, we propose a quantum algorithm to obtain the lowest energy eigenstate of an arbitrary Hamiltonian simulated by a quantum computer. Similarly to some AQC methods \cite{Boykin02,Xu14} (or other variational methods in classical computation \cite{Kosloff86,Lehtovaara07}), we start from an {\em arbitrary} initial state of the simulated system that (usually) contains very small fractions of the lower energy eigenstates. Applying a finite series of transformations, each of which consists of the quantum Householder reflection and the unitary of the system dynamics, the initial state is allowed to evolve amplifying the fraction of the lowest energy eigenstate up to $\simeq 1$. We note that our algorithm also employs an ancillary qubit system, similarly to DLAC method. However, no measurements are performed on the ancillar-qubit during the algorithm process; namely, {\em Demon is not needed in our algorithm}. The measurement is performed only once at the end of the algorithm to get the final state of the system removing the ancillary qubit \footnote{We can also observe the cooling-like and the heating-like behaviors when we consider the whole system Hamiltonian involving the ancillary qubit system (See Appendix B for details).}. In our theoretical analysis, we prove that our algorithm faithfully works for any given Hamiltonian. Numerical analyses are also carried out. Firstly, we provide a numerical proof-of-principle demonstration for a simple Hamiltonian whose eigenvalues are equally spaced. The result is quite consistent with our theoretical analysis, and in particular it exhibits the behavior comparable to the best `cooling' available with the DLAC method. We then consider a random Hamiltonian model for further analysis of our algorithm. We presume that the required iterations to achieve an accuracy $1-\epsilon$ ($\epsilon \ll 1$) is proportional to $\simeq {\cal O}(D^{-\alpha}\epsilon^{-\beta})$ with $\alpha, \beta \le 1$, where $D$ is difference of the two lowest eigenvalues, and $\epsilon$ is tolerable error defined as the probability that the finally obtained state of the system is to be unexpected (i.e. not the lowest) eigenstate. By numerical simulations, it is found that $\alpha \simeq 1$ and $\beta \simeq 0.19 < 1$.

%----------------------------
\section{Problem \& Method}\label{sec:2}

To begin, we state the problem as follows: Consider a Hamiltonian $\hat{H}$ of $N$-dimensional Hilbert-space. The energy eigenvalues $\lambda_k$ of $\hat{H}$ are scaled as
\begin{eqnarray}
0 < \lambda_0 < \lambda_1 \le \ldots \le \lambda_{N-1} \le 1,
\label{eq:normalized_eigen}
\end{eqnarray}
where we assume no perfect degeneracy between $\lambda_0$ and $\lambda_1$ (i.e. $\lambda_1 - \lambda_0 \neq 0$). The energy eigenstate associated with the energy eigenvalue $\lambda_k$ is given as $\ket{\lambda_k}$. Note that the energy eigenvalues $\lambda_k$ and the eigenstates $\ket{\lambda_k}$ are {\em completely unknown}. In this circumstance, the problem that we focus on here is: How to obtain the lowest energy eigenstate $\ket{\lambda_0}$. 
%This is relevant to the problem of ``pseudo-cooling'', even when the temperature could not be well-defined in the process \cite{Xu14}.

In solving this problem, we start with an initial system state $\ket{\varphi_0}$ in $N$-dimensional Hilbert-space of the given Hamiltonian,
\begin{eqnarray}
\ket{\varphi_0} = \sum_{j=0}^{N-1}a_j\ket{v_j},
\end{eqnarray}
where the computational bases $\ket{v_j}$ and the coefficients $a_j$ are {\em known} to us. We assume that the fraction $f_0(\lambda_0) =\abs{\braket{\lambda_0}{\varphi_0}}^2$ is not equal to zero but vanishingly small. We then consider an ancillary system of a clean qubit. By adopting an ancilla-qubit state $\ket{\phi_0}=\frac{1}{\sqrt{2}}(\ket{0}+\ket{1})$, we prepare a composite initial state $\ket{\psi_0}$ such that
\begin{eqnarray}
\ket{\psi_0} = \ket{\phi_0} \otimes \ket{\varphi_0} =\frac{\ket{0}+\ket{1}}{\sqrt{2}} \otimes \sum_{j=0}^{N-1}a_j\ket{v_j}.
\end{eqnarray}
Here, if we expand $\ket{\varphi_0}$ in terms of the eigenstates $\ket{\lambda_k}$, the composite initial state $\ket{\psi_0}$ is rewritten as
\begin{eqnarray}
\ket{\psi_0} = \sum_{k=0}^{N-1}\frac{\gamma_{0,k}}{\sqrt{2}}\Big(\ket{0,\lambda_{k}}+\ket{1,\lambda_{k}}\Big),
\label{eq:init_st}
\end{eqnarray}
where $\ket{0,\lambda_{k}}=\ket{0}\otimes\ket{\lambda_k}$, $\ket{1,\lambda_{k}}=\ket{1}\otimes\ket{\lambda_k}$, and ${\gamma_{0,k}=\sum_{j=0}^{N-1}a_j\braket{\lambda_k}{v_j}}$. Here, $|\gamma_{0,0}|^2 = f_0(\lambda_0)=\abs{\braket{\lambda_0}{\varphi_0}}^2$. Note that the coefficients $\gamma_{0,k}$ cannot be evaluated, because the energy eigenstates $\ket{\lambda_k}$ are unknown. 

Then, we construct a finite number $n_c$ of transformations, $\hat{T}_1, \hat{T}_2, \ldots, \hat{T}_{n_c}$, according to the following {\em recursive} relation: For $i=1,2,\ldots,n_c$,
\begin{eqnarray}
\hat{R}_{i} = \hat{T}_{i}\hat{R}_{i-1}\hat{T}_{i}^{\dagger},~\text{and}~\hat{T}_{i} = \hat{R}_{i-1}\hat{U}\hat{R}_{i-1}\hat{U}^{\dagger}, 
\label{eq:t_recur}
\end{eqnarray}
where $\hat{R}_0$ is ``quantum Householder reflection'', defined as $\hat{R}_0 = \hat{\identity} - 2\ket{\psi_0}\bra{\psi_0}$. Such an operation has widely been used in quantum search \cite{Grover97}, or other tasks \cite{Ivanov06,Ivanov07,Ivanov08}. We also use a $2N$-dimensional unitary $\hat{U}$, defined as
\begin{eqnarray}
\hat{U}=\left(\ket{0}\bra{0}\otimes\hat{A}(\tau)\right) + i\left(\ket{1}\bra{1}\otimes \hat{A}(\tau)^\dagger\right),
\label{eq:uop}
\end{eqnarray}
where $\hat{A}(\tau)=e^{i \frac{\pi}{4}\tau \hat{H}}$ is a unitary of the system's dynamics \footnote{Here, we omit the conventional minus (`$-$') sign in the exponent.}, and $\tau \in (0, 1]$ is a scaling constant concerning the time of Hamiltonian action.

\begin{figure}[t]
\centering
\includegraphics[width=0.5\textwidth]{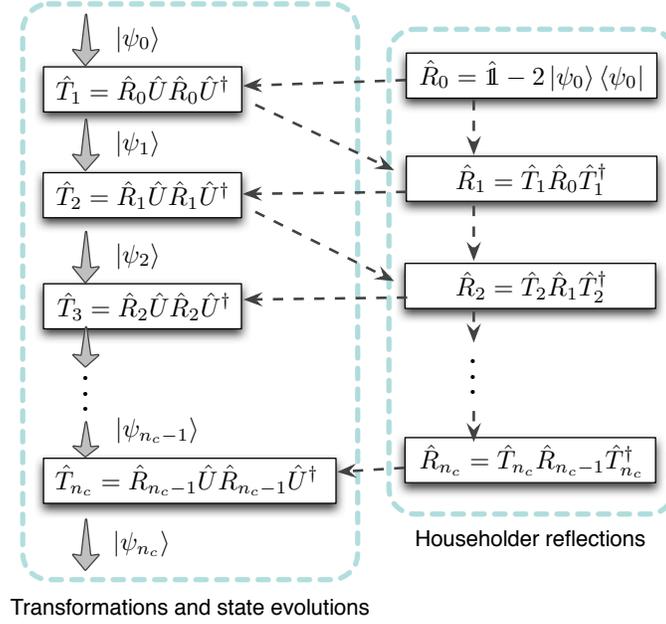}
\caption{(Color online) Schematic picture of the process of our algorithm (See the main text)}
\label{fig:process}
\end{figure}

Then, the process of our algorithm can simply be thought of as a series of transformations $\hat{T}_i$ applied on the initial composite state $\ket{\psi_0}$ such that (See Fig.~\ref{fig:process}):
\begin{eqnarray}
\hat{T}_{n_c}\ldots\hat{T}_2\hat{T}_1\ket{\psi_0}=\ket{\psi_{n_c}}.
\label{eq:alg_process}
\end{eqnarray}
At the final step of $n_c$, measurement is performed only once on the ancillary qubit. Tracing out the ancilar-qubit state by the measurement, the remaining state of the system is supposed to be close to the lowest energy eigenstate $\ket{\lambda_0}$.

%----------------------------
\section{Theoretical analysis}\label{sec:3}

We now analyze the process of our algorithm. By using Eqs.~(\ref{eq:init_st})-(\ref{eq:alg_process}), we can describe the evolution of the composite state (See Appendix A):
\begin{eqnarray}
\ket{\psi_0} &\overset{\hat{T}_1}{\longrightarrow}& \ket{\psi_1} = \sum_{k=0}^{N-1}\frac{\gamma_{0,k}}{\sqrt{2}}\Big(\gamma_{1,k}\ket{0,\lambda_{k}} + \gamma_{1,k}^\ast\ket{1,\lambda_{k}}\Big) \nonumber \\
&\overset{\hat{T}_2}{\longrightarrow}& \ket{\psi_2}=\sum_{k=0}^{N-1}\frac{\gamma_{0,k}}{\sqrt{2}}\Big(\gamma_{1,k}\gamma_{2,k}\ket{0,\lambda_{k}} + \gamma_{1,k}^\ast\gamma_{2,k}^\ast\ket{1,\lambda_{k}}\Big) \nonumber \\
		&\vdots& \nonumber \\
&\overset{\hat{T}_{i}}{\longrightarrow}& \ket{\psi_{i}}=\sum_{k=0}^{N-1}\frac{\gamma_{0,k}}{\sqrt{2}}\Big(\Gamma_{i,k}\ket{0,\lambda_{k}} + \Gamma_{i,k}^\ast\ket{1,\lambda_{k}}\Big) \nonumber \\
&\vdots&
%&\overset{\hat{T}_{i}}{\longrightarrow}& \ket{\psi_i}=\sum_{k=0}^{N-1}\frac{\gamma_{0,k}}{\sqrt{2}}\left(\prod_{l=1}^{i}\gamma_{l,k}\ket{0,\lambda_{k}} + \prod_{l=1}^{i}\gamma_{l,k}^\ast\ket{1,\lambda_{k}}\right).
\label{eq:st_evolve}
\end{eqnarray}
where $\Gamma_{i,k} = \gamma_{1,k}\gamma_{2,k}\cdots\gamma_{i,k}$ (thus, $\Gamma_{i,k}^\ast=\gamma_{1,k}^\ast\gamma_{2,k}^\ast\cdots\gamma_{i,k}^\ast$), and $\gamma_{i,k}$ is given as a function of $k$th-ordered energy eigenvalue $\lambda_k$ (Here, $\Gamma_{0,k}=1$). The explicit form of $\gamma_{i,k}$ is given as
\begin{eqnarray}
\gamma_{i,k} = \left(4\abs{W_{i-1}}^2-1\right) - 2\abs{W_{i-1}}e^{-i \frac{\pi \left(1 -\tau\lambda_k\right)}{4}}.
\label{eq:gamma}
\end{eqnarray}
where the factor $\abs{W_i}$ is defined as
\begin{eqnarray}
\abs{W_i} = \abs{\bra{\psi_{i}}\hat{U}\ket{\psi_{i}}} = \sum_{k=0}^{N-1}\abs{\gamma_{0,k}}^2\abs{\Gamma_{i,k}}^2\cos{\frac{\pi \left(1 - \tau\lambda_k\right)}{4}}.
\label{eq:factor_Wi}
\end{eqnarray}

Then, after sufficiently large $n_c \gg 1$ step, we perform a von-Neumann measurement with $\{\ket{0}, \ket{1}\}$ on the ancillary qubit. The ancillary qubit is then removed, and we get the final state $\ket{\varphi_{n_c}}$ involving the fractions $f_{n_c}(\lambda_k)=\abs{\gamma_{0,k}}^2\abs{\Gamma_{n_c,k}}^2$ of the energy eigenstates. More specifically, the final state $\ket{\varphi_{n_c}}$ is given as follows (Note that $\sum_{k=0}^{N-1}\abs{\gamma_{0,k}}^2\abs{\Gamma_{i,k}}^2 = 1$).
\begin{eqnarray}
\ket{\varphi_{n_c}}\to
\left\{
\begin{array}{ll}
\displaystyle{\sum_{i=0}^{N-1}\gamma_{0,k}\Gamma_{n_c,k}\ket{\lambda_k}} & ~\text{if $\ket{0}$ is measured (with $\frac{1}{2}$ probability)}, \\
\displaystyle{\sum_{i=0}^{N-1}\gamma_{0,k}\Gamma_{n_c,k}^\ast\ket{\lambda_k}} & ~\text{if $\ket{1}$ is measured (with $\frac{1}{2}$ probability)},
\end{array}
\right.
\label{eq:final_state}
\end{eqnarray}

We now show that the final state $\ket{\varphi_{n_c}}$ in Eq.~(\ref{eq:final_state}), either $\sum_{i=0}^{N-1}\gamma_{0,k}\Gamma_{n_c,k}\ket{\lambda_k}$ or $\sum_{i=0}^{N-1}\gamma_{0,k}\Gamma_{n_c,k}^\ast\ket{\lambda_k}$, becomes close to $\ket{\lambda_0}$ in the limit of $n_c \to \infty$. In particular, we will show that
\begin{eqnarray}
f_{0}(\lambda_0) \le f_{1}(\lambda_0) \le \cdots < f_{n_c-1}(\lambda_0) \le f_{n_c}(\lambda_0) \to 1, ~(\text{as}~n_c \to \infty).
\label{eq:p0_n}
\end{eqnarray}
To this end, we state and sketch the proof of the following two propositions: [{\bf P.1}] The coefficient factor $\abs{\gamma_{i,0}}^2$ of the lowest eigenstate is {\em always} larger than or equal to $1$, and thus, $f_i (\lambda_0) \le f_{i+1} (\lambda_0)$ for all $i = 0,1,\ldots, \infty$. [{\bf P.2}] The other probabilities $f_i(\lambda_{k \neq 0})$ of the higher eigenstates (excepting the lowest one) goes to zero with $\abs{\gamma_{i,{k \neq 0}}}^2 < 1$ when $i \to \infty$. The proof of the latter is particularly important to guarantee that $f_i(\lambda_0)$ can go {\em up to $\simeq 1$}.

First, we provide the proof of [{\bf P.1}]. To start, let us calculate $\abs{\gamma_{i,k}}^2$ from Eq.~(\ref{eq:gamma}). With the simple algebra, we have
\begin{eqnarray}
\abs{\gamma_{i,k}}^2 = 1 + 4\abs{W_{i-1}} \left( 4\abs{W_{i-1}}^2 -1\right) \Delta_{i-1,k},
\label{eq:Q_ik}
\end{eqnarray}
where 
\begin{eqnarray}
\Delta_{i-1,k} = \abs{W_{i-1}} - \cos \frac{\pi \left(1-\tau\lambda_k\right)}{4}. 
\label{eq:delta}
\end{eqnarray}
By observing Eq.~(\ref{eq:Q_ik}), we can know that the proof of [{\bf P.1}], i.e. $\abs{\gamma_{i,0}}^2 \ge 1$, is straightforward if 
\begin{eqnarray}
\left\{
\begin{array}{l}
(a) \abs{W_{i-1}}^2 \ge \frac{1}{2}~(\forall i), \\ 
(b) ~\Delta_{i-1,0} \ge 0 ~(\forall i).
\end{array} 
\right.
\label{eq:condi_ab}
\end{eqnarray}
Note that $\abs{W_{i-1}}$ is larger than $0$. To verify Eq.~(\ref{eq:condi_ab}), we give the {\em theoretical} lower and upper bound on the value of $\abs{W_{i-1}}$, by using Eq.~(\ref{eq:factor_Wi}), as
\begin{eqnarray}
\cos{\frac{\pi \left(1-\tau\lambda_0\right)}{4}} \le \abs{W_{i-1}}=\left< \cos{\frac{\pi \left(1-\tau\lambda_k\right)}{4}} \right>_{i-1} \le \cos{\frac{\pi \left(1-\tau\lambda_{N-1}\right)}{4}},
\label{eq:bound_wi}
\end{eqnarray}
where `$\left< x_k \right>_{i-1} = \sum_{k=0}^{N-1} f_{i-1}(\lambda_k)x_k$', which is an expectation value of $x_k$ at ($i-1$)th step. Here, the (mathematical) condition to meet the lower bound is that $f_{i-1}(\lambda_0)=1$ and $f_{i-1}(\lambda_{k\neq 0})=0$, whereas the upper bound is given when $f_{i-1}(\lambda_{N-1})=1$ and $f_{i-1}(\lambda_{k \neq N-1})=0$. Note that Eq.~(\ref{eq:bound_wi}) is {\em always satisfied} for the given eigenvalues $\lambda_k$ ($k=0,1,\ldots,N-1$) scaled as in Eq.~(\ref{eq:normalized_eigen}). Thus, ($a$) $\abs{W_{i-1}}^2  \ge \frac{1}{2}$ ($\forall i$) is true (because $\lambda_0 > 0$). Then, by applying the lower bound value `$\cos{\frac{\pi \left(1-\tau\lambda_0\right)}{4}}$' to Eq.~(\ref{eq:delta}) with $k=0$, we can directly verify that ($b$) $\Delta_{i,0} \ge 0$ ($\forall i$) also holds. Therefore, [{\bf P.1}] is always the case. 

Next, let us consider [{\bf P.2}]. To proceed, we assume that the eigenvalues $\lambda_k$ ($k=0,1,\ldots,N-1$) are divided into the two groups $g_1$ and $g_2$ at any ($i-1$)th step, each of which is characterized by 
\begin{eqnarray}
\left\{
\begin{array}{ll}
g_1 : & f_{i-1}(\lambda_k) \le f_{i}(\lambda_k)~\text{for}~\lambda_k \le \xi_{i-1}, \\
g_2 : & f_{i-1}(\lambda_k) > f_{i}(\lambda_k)~\text{for}~\lambda_k > \xi_{i-1}, 
\end{array}
\right.
\end{eqnarray}
where $\xi_{i-1}$ is a boundary factor of the ($i-1$)th step. Then, by using Eqs.~(\ref{eq:Q_ik})-(\ref{eq:condi_ab}), we can show that
\begin{eqnarray}
\abs{\gamma_{i,k}}^2 \ge 1 \Longleftrightarrow \Delta_{i-1,k} \ge 0 \Longleftrightarrow \lambda_k \le \xi_{i-1},
\end{eqnarray}
where the explicit form of the boundary factor $\xi_{i-1}$ can be found as
\begin{eqnarray}
\xi_{i-1} = \frac{1}{\tau}\left( 1 - \frac{4}{\pi}\arccos{\abs{W_{i-1}}}\right).
\label{eq:xi}
\end{eqnarray}
Here, noting [{\bf P.1}] and the property $\sum_{k=0}^{N-1}f_i(\lambda_k)=\sum_{k=0}^{N-1}f_{i-1}(\lambda_k)=1$, we can verify that $\abs{W_{i-1}} \ge \abs{W_{i}}$ ($\forall i$), since the increment of $f_{i-1}(\lambda_0)$ necessarily results in the decrements of any other probabilities $f_{i-1}(\lambda_{k \neq 0})$. This allows us to prove that [by using Eq.~(\ref{eq:xi})]
\begin{eqnarray}
\xi_{i} \le \xi_{i-1} ~\text{for all}~i=1,2,\ldots,\infty.
\label{eq:in_xi}
\end{eqnarray} 
By using Eqs.~(\ref{eq:bound_wi}) and (\ref{eq:xi}), we can give the {\em theoretical} lower and upper bound of the boundary factor $\xi_{i}$, for any $i$, as 
\begin{eqnarray}
\lambda_0 \le \xi_{i} \le \lambda_{N-1}.
\label{eq:lb_xi}
\end{eqnarray}
We note that the theoretical lower bound in Eq.~(\ref{eq:lb_xi}) is an alternative expression of [{\bf P.1}]; namely, it is true that the lowest eigenvalue $\lambda_0$ {\em always} belongs to the group $g1$. By using Eq.~(\ref{eq:xi}), we can also obtain that \footnote{Note that $\arccos{(x)} = \frac{\pi}{2} - \arcsin{(x)}$, and $\arcsin{(x)} = \sum_{l=0}^{\infty} \frac{(2l)!}{4^l (2l+1) (l!)^2} x^{2l+1}$ ($\abs{x} \le 1$).}
\begin{eqnarray}
\xi_{i-1} - \xi_{i}  &=& \frac{4}{\tau\pi}\Big( \arccos{\abs{W_{i}}}-\arccos{\abs{W_{i-1}}} \Big) \nonumber \\
    &=& \frac{4}{\tau\pi}\sum_{l=0}^{\infty} \frac{(2l)!}{4^l (2l+1) (l!)^2}\left( \abs{W_{i-1}}^{2l+1} - \abs{W_{i}}^{2l+1} \right).
%    &>& \frac{4}{\tau\pi}\left( \abs{W_{i-1}}-\abs{W_{i}}\right).
\label{eq:d_xi}
\end{eqnarray}
Here, (mathematically) if $\abs{W_{i-1}} - \abs{W_i} \not\to 0$, then $\xi_{i-1} - \xi_i \not\to 0$, i.e. $\xi_{i}$ does not converge but becomes smaller as increasing $i \to \infty$ [as in Eq.~(\ref{eq:in_xi})]. However, physically, $\xi_{i}$ must converge to a finite value {\em larger than} or {\em equal to} $\lambda_0$ with $\abs{W_{i-1}} - \abs{W_{i}} \to 0$ when $i \to \infty$. Thus, we assume that $\xi_{i}$ is converged to a value in between $\lambda_{k'}$ and $\lambda_{k'+1}$, where $k'$ is a specific (integer) number of the eigenvalue index. From this, we assume further that, for $i \to \infty$,
\begin{eqnarray}
f_i(\lambda_{k \le k'}) \to u_k ~\text{and}~ f_i(\lambda_{k > k'}) \to 0,
\label{eq:assume_s}
\end{eqnarray} 
where $\sum_{k \le k'} u_k = 1$. Then, by using Eq.~(\ref{eq:factor_Wi}), we obtain that
\begin{eqnarray}
\abs{W_{i-1}} - \abs{W_{i}} \to \sum_{k \le k'} \left(1-\abs{\gamma_{{i},k}}^2\right)u_k\cos\frac{\pi \left(1-\tau\lambda_k\right)}{4}, ~\text{for}~i \to \infty.
\label{eq:wi_q}
\end{eqnarray}
In the circumstance, we can {\em infer} that the only possible solution for the physically reasonable condition (i.e. $\abs{W_{i-1}} - \abs{W_{i}} \to 0$) is given by
\begin{eqnarray}
\abs{\gamma_{i,0}}^2 \to 1 ~\&~ \abs{\gamma_{i,k \neq 0 \le k'}}^2 < 1, ~\text{and},~ u_{k \neq 0 \le k'} = 0,
\label{eq:sol2}
\end{eqnarray}
when $i \to \infty$. This solution yields that (for $i \to \infty$),
\begin{eqnarray}
\left\{
\begin{array}{l}
f_i(\lambda_0) \to u_0 = 1 ~(\text{because}~\sum_{k \le k'} u_k = 1), \\
\abs{W_{i}} \to \cos\frac{\pi \left(1-\tau\lambda_0\right)}{4}, ~\text{and thus,}~ \xi_{i} \to \lambda_0 ~[\text{from Eqs.~(\ref{eq:factor_Wi}), and (\ref{eq:xi})}], 
\end{array}
\right.
\label{eq:sol_properties}
\end{eqnarray}
which are consistent with Eq.~(\ref{eq:sol2}) again \footnote{i.e. from $\abs{W_i} \to \cos\frac{\pi \left(1-\tau\lambda_0\right)}{4}$, it is verified that $\Delta_{i,0}\to 0$ [from Eq.~(\ref{eq:delta})], and thus, $\abs{\gamma_{i+1,0}}^2 \to 1$ [as in Eq.~(\ref{eq:sol2})].}. Here, if we assume any {\em nonzero} value(s) of $u_{k \neq 0 \le k'}$ with $\abs{\gamma_{i,k \neq 0 \le k'}}^2 \to 1$ in Eq.~(\ref{eq:assume_s}) (and hence, $u_0 < 1$), it is in contradiction to the assumption (For detailed proof, see Appendix B). On the basis of the above description, it is certain that all the probabilities $f_i(\lambda_{k\neq 0})$ of the higher eigenstates, excepting the lowest one, go to zero [i.e. $f_i(\lambda_{k > k'}) \to 0$ from Eq.~(\ref{eq:assume_s})~\&~$f_i(\lambda_{k \neq 0 \le k'}) \to 0$ from Eq.~(\ref{eq:sol2})] when $i \to \infty$. 
%Thus, for physically reasonable condition (i.e. $\abs{W_{i-1}} - \abs{W_{i}} \to 0$), we should consider any coefficient factor(s) $\abs{\gamma_{i,k \neq 0}}^2 < 1$ ($k < k'$) as long as [{\bf P.1}] (i.e. $\abs{\gamma_{i,0}}^2 > 1$) is true [See Eq.~(\ref{eq:wi_q})], however, it would not yield the nonzero value(s) $u_{k \neq 0}$ when $i \to \infty$. 

Therefore, we can show that the fraction $f_i(\lambda_0)$ of the lowest energy eigenstate only reaches close to unity after a sufficiently large number $n_c$ of the iterations [as in Eq.~(\ref{eq:p0_n})], whereas all other probabilities $f(\lambda_{k \neq 0})$ go to zero.

%----------------------------
\section{Numerical analysis}\label{sec:4}

\begin{figure}[t]
\centering
\includegraphics[angle=270,width=0.35\textwidth]{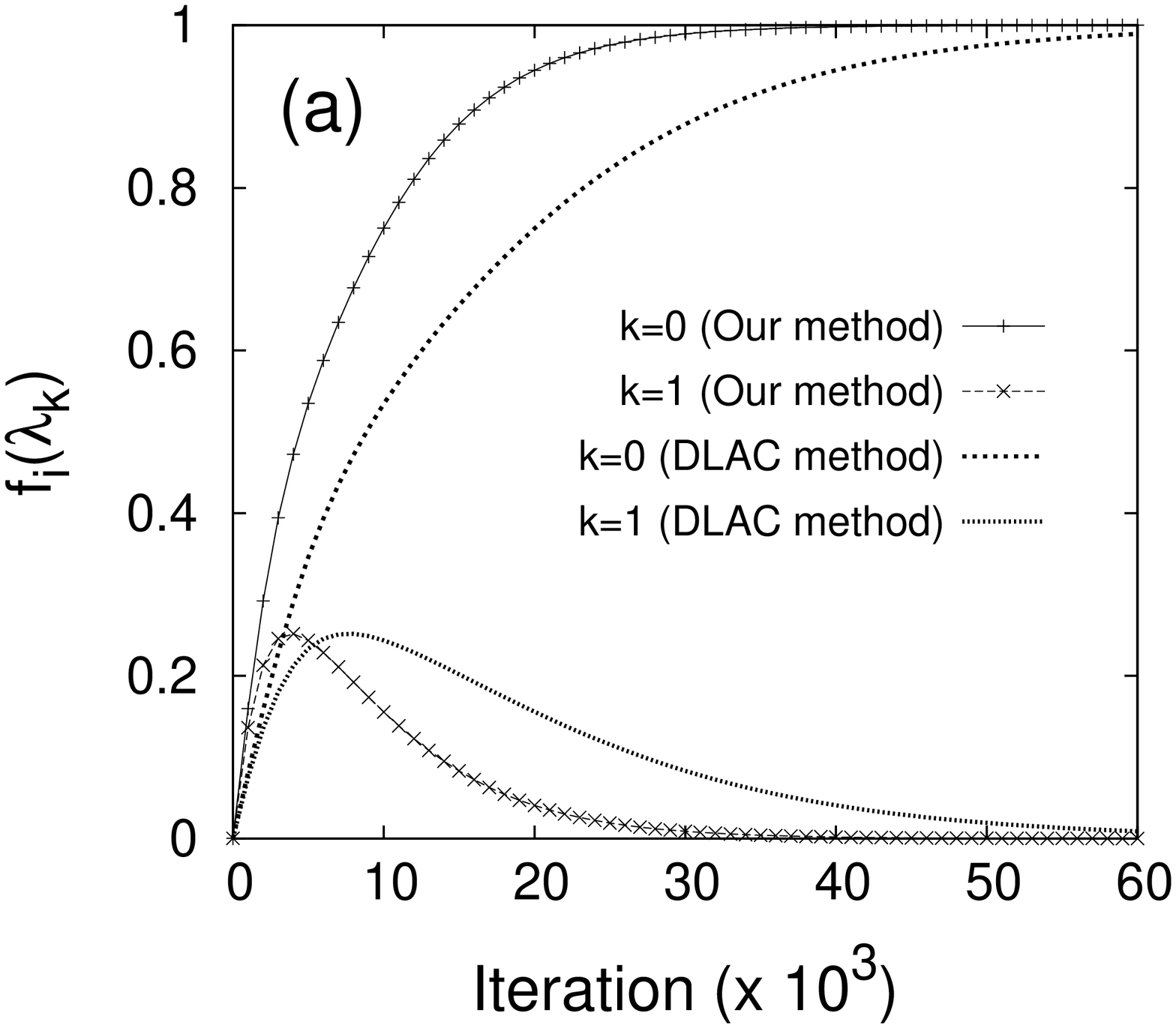}
\includegraphics[angle=270,width=0.35\textwidth]{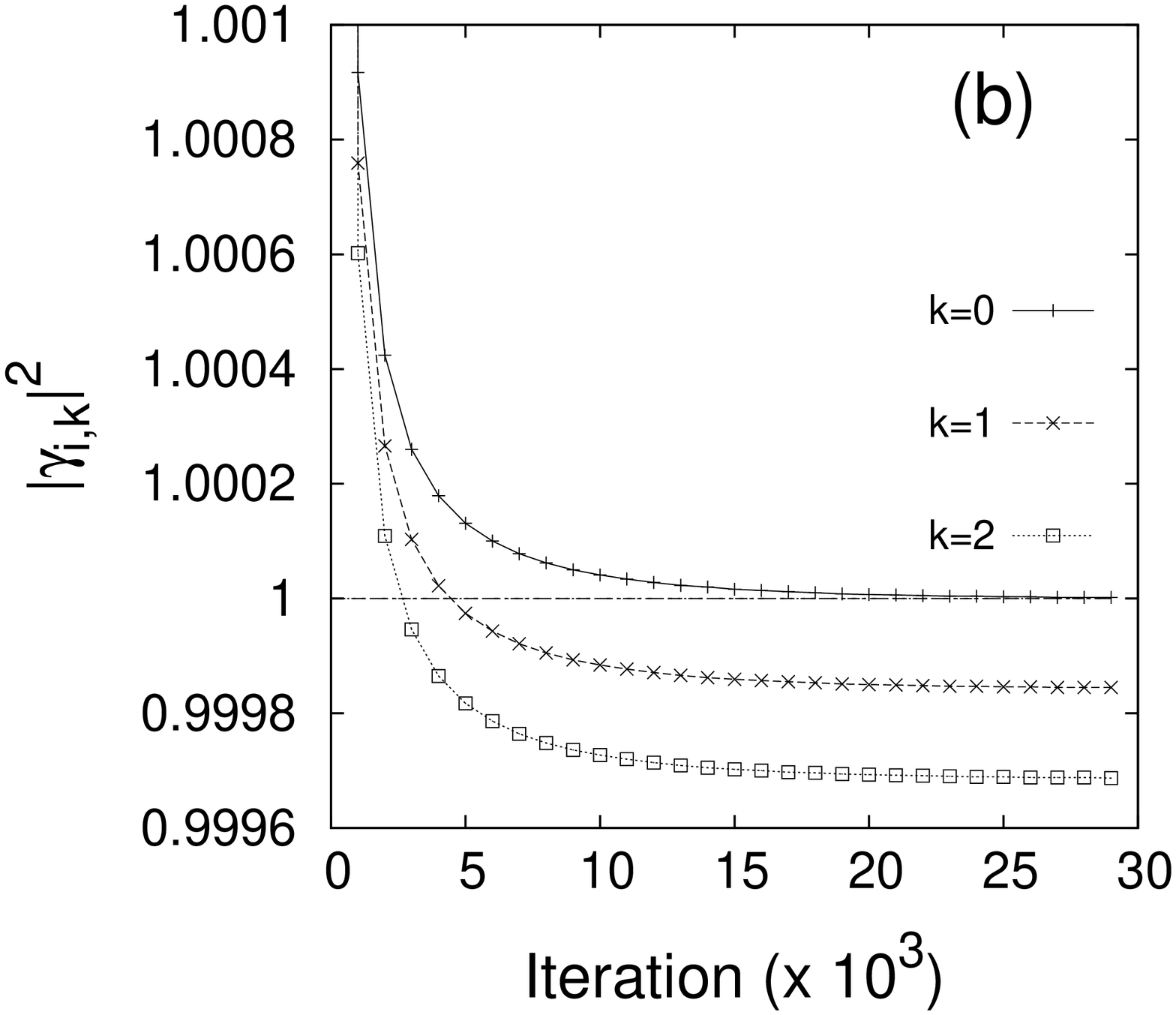}
\caption{(a) We plot $f_i(\lambda_0)$ and $f_i(\lambda_1)$ for $N=10^4$. Here we chose a specific $\hat{H}$ whose eigenvalues are equally spaced from $E_0 \le \frac{1}{N}$ to $E_0 + \frac{N-1}{N}$. In the simulation, the initial system state $\ket{\varphi_0}$ is chosen with $\gamma_{0,k}=\frac{1}{\sqrt{N}}$ for simplicity. We can see that $f_i(\lambda_0)$ reaches from $10^{-4}$ to $1$, and $f_i(\lambda_1)$ decay down to $0$. For comparison, we also draw $f_i(\lambda_0)$ and $f_i(\lambda_1)$, assuming that we use the DLAC method [See Eq.~(\ref{eq:cfactor_DLAC})] and the same initial system state $\ket{\varphi_0}$ is used. (b) The graphs of $\abs{\gamma_{i,k}}^2$ are given for $k=0,1,2$. Here, $\abs{\gamma_{i,0}}^2 > 1$ for all $i$. These are consistent with our analyses.}
\label{grp:probs}
\end{figure}

We firstly provide a numerical proof-of-principle demonstration for a simple physical system where the energy levels are equally spaced (sometimes, called ``Wannier-Stark ladder'' \cite{Wannier60,Mendez88,Voisin88}). We thus consider a $N$-dimensional Hamiltonian $\hat{H}$ with the equidistant eigenvalues $\lambda_k = E_0 + \frac{k}{N}$ ($k=0,1,\ldots, N-1$). Here, the ground-state energy $E_0$ has a finite value less than $\frac{1}{N}$. In the simulation, we set the dimension of the Hilbert-space as $N=10^4$, and, for simplicity, the initial system state $\ket{\varphi_0}$ is chosen with $\gamma_{0,k}=\frac{1}{\sqrt{N}}$ for all $k$. Note that the chosen state $\ket{\varphi_0}$ contains a very small fraction of the ground state, i.e. $f_0(\lambda_0) = 10^{-4}$. In Fig.~\ref{grp:probs}(a), we plot $f_i(\lambda_0)$ and $f_i(\lambda_1)$ with increasing the iteration $i$, where $f_i({\lambda_0})$ reaches from $10^{-4}$ to close to $1$, and $f_i(\lambda_{1})$ decays down to $0$. For a comparison, we also include the data of $f_i(\lambda_0)$ and $f_i(\lambda_1)$ in Fig.~\ref{grp:probs}(a), assuming that we use the DLAC method. Here we consider, particularly, the best `cooling' available with the DLAC method (i.e. the case where the cooled results are only appeared in the Demon's measurements). We note that, in DLAC method, the `cooling' factor (similar to the factor $\gamma_{i,k}$ in our algorithm) is given, for the system dynamics $\hat{A}(\tau) = e^{-i \frac{\pi}{4}\hat{H} \tau}$, as \cite{Xu14}
\begin{eqnarray}
1-\sin{\xi_k} \approx e^{-\frac{\pi}{4}\lambda_k \tau},
\label{eq:cfactor_DLAC}
\end{eqnarray}
where $\xi_k \in [-\frac{\pi}{2}, \frac{\pi}{2}]$ is associated with the energy eigenvalues $\lambda_k$, and the approximation is done for small time evolution \footnote{This `cooling' behavior is also similar to that of a {\em classical} method, called ``imaginary time propagation (ITP)'' \cite{Kosloff86,Lehtovaara07}. Actually, if we consider that the given Hamiltonian is diagonalized, we can associate these two methods (See Supplementary Information of Ref.~\cite{Xu14}).}. Thus, we adopt the same initial state $\ket{\varphi_0}$ to make the comparison as convincing as possible. It is observed that, in this case, $f_0(\lambda_0)$ also grows up to $1$, and the behavior is quite similar to that of our algorithm. We also plot the coefficients $\abs{\gamma_{i,0}}^2$ and $\abs{\gamma_{i,1}}^2$ in Fig.~\ref{grp:probs}(b), where $\abs{\gamma_{i,0}}^2$ is always larger than $1$, whereas $\abs{\gamma_{i,1}}^2$ becomes smaller than $1$ leading to the decrease of $f_i(\lambda_{1})$. These are consistent with the above theoretical analyses in Eq.~(\ref{eq:sol2}). 

From the analyzed behaviors in the previous section and already known classical methods \cite{Abrams99,Yung12}, we presume here that the number $n_c$ of iterations for an accuracy $f_{n_c}(\lambda_0) > 1-\epsilon$ ($\epsilon \ll 1$) is dominated by the difference of the two lowest eigenvalues, $D=\lambda_1 - \lambda_0 \ll 1$, and the tolerable error $\epsilon = 1-f_{n_c}(\lambda_0)$. If we assume that $\lambda_0$ and $\lambda_1$ are so small that all of the higher-order terms (e.g. $\lambda_{0}^2, \lambda_{0}^3, \ldots$, and $\lambda_{1}^2, \lambda_{1}^3, \ldots$) can be negligible, we can explicitly calculate that $n_c$ is upper bounded as $n_c \le \frac{2}{\pi} \tau^{-1} D^{-1} \epsilon^{-1}$. Thus, more generally, we conjecture that $n_c \simeq {\cal O}(cD^{-\alpha} \epsilon^{-\beta})$ with $\alpha, \beta \le 1$. Here, $c$ is a constant factor. 

\begin{figure}[t]
\centering
\includegraphics[angle=270,width=0.35\textwidth]{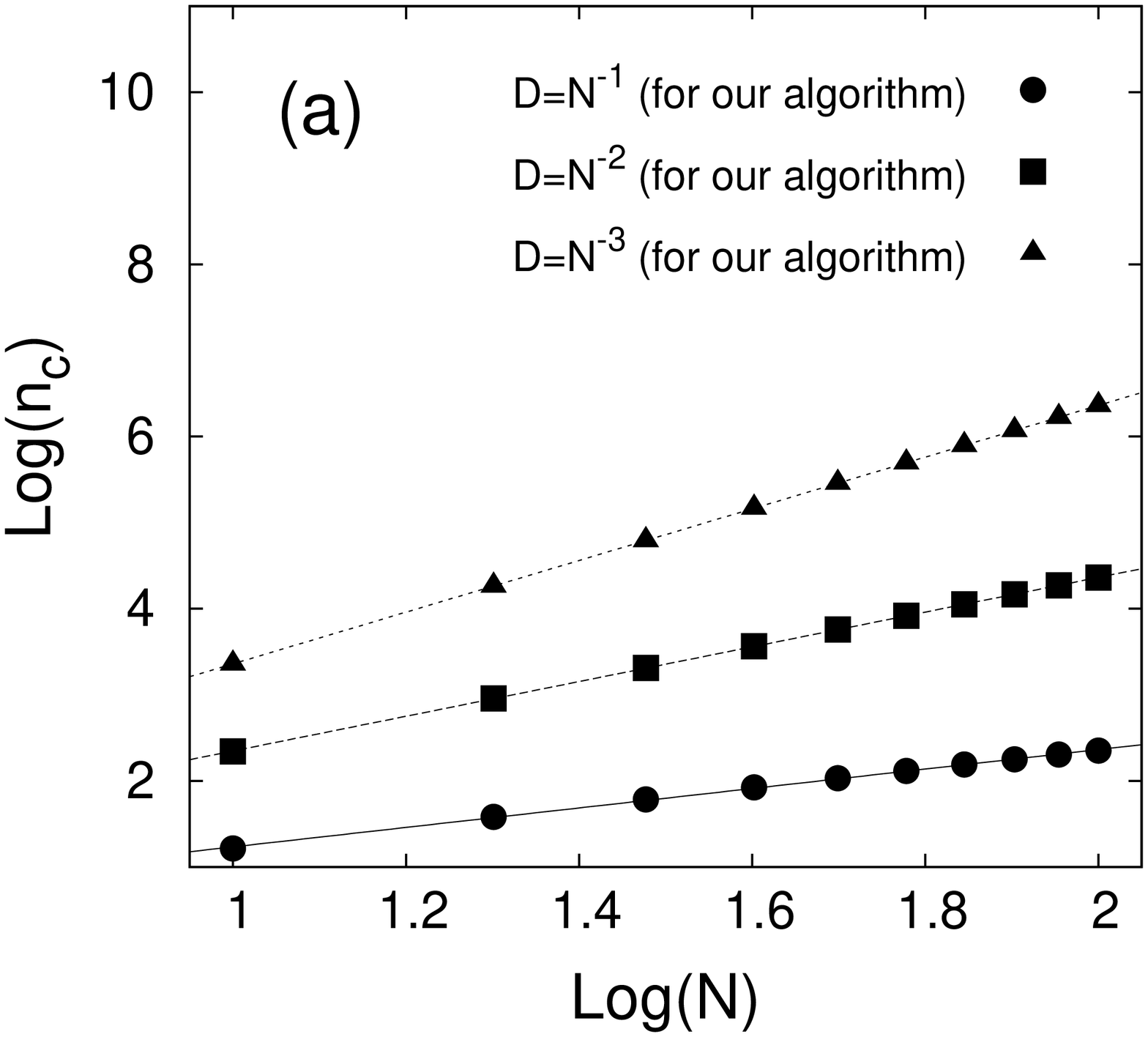}
\includegraphics[angle=270,width=0.35\textwidth]{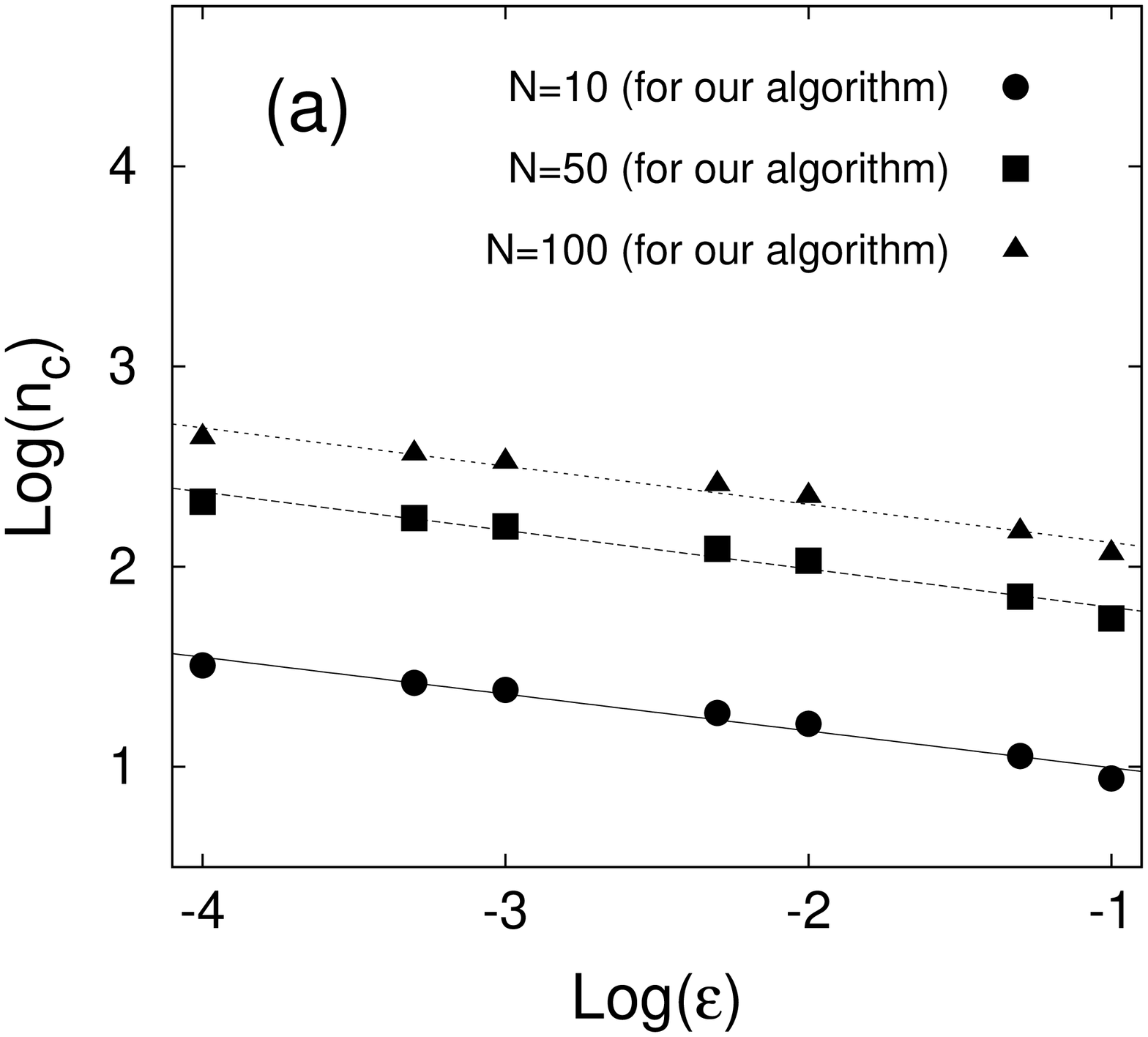}
\caption{(a) We give $n_c$ versus $N$ graphs on a log-log scale for the random Hamiltonian. The iteration is terminated when $f_i(\lambda_0) \ge 0.99$, i.e. $\epsilon \le 0.01$. We consider three different cases: $D=1/N$, $1/N^{2}$, and $1/N^{3}$, where $D=\lambda_1 - \lambda_0$. Each data point is made by averaging $1000$ simulations. The data are well fitted to $\log{n_c}=A\log{N}+B$ for $D=1/N$ (solid line), $D=1/N^2$ (dashed line), and $D=1/N^3$ (dotted line). (b) We also give $n_c$ versus $\epsilon$ graphs on a log-log scale. In this case, $D$ is fixed to $1/N$, and we consider three cases by taking $N=10$, $50$, and $100$. Each data is also averaged over $1000$ simulations. The Hamiltonian is also randomly generated in each simulation. The data are well fitted to $\log{n_c}=A\log{\epsilon}+B$ for $N=10$ (solid line), $N=50$ (dashed line), and $N=100$ (dotted line). (See the main text for detailed fitting parameters)}
\label{grp:rand_anal}
\end{figure}

With the above prediction in mind, we consider a model of randomly generated Hamiltonian for more general analysis. We perform numerical simulations, and find $n_c$ for a given accuracy level. In the simulations, number $N$ of energy eigenvalues are randomly generated \footnote{Here, we construct the random Hamiltonian in such a way: Firstly, we make a diagonal matrix $\hat{d}=\text{diag}\{\lambda_0, \lambda_1, \ldots, \lambda_{N-1}\}$ with a randomly generated energy eigenvalue $\lambda_k$ (but, $D = \lambda_1 - \lambda_0$ is always be a certain predetermined value). We, then, construct an Hamiltonian by rotating $\hat{d}$ such that $\hat{V} \hat{d} \hat{V}^\dagger$, where the unitary $\hat{V}=e^{-i \mathbf{p}\cdot\mathbf{G}}$ is given by the randomly chosen real parameter vector $\mathbf{p}=(p_1, p_2, \ldots, p_{N^2-1})^T$ and SU($N$) group generators $\mathbf{G}=(\hat{g}_1, \hat{g}_2, \ldots, \hat{g}_{N^2-1})^T$ \cite{Hioe81}.} in ($0, 1$], but $\lambda_1$ and $\lambda_0$ are chosen such that the difference $D = \lambda_1 - \lambda_0$ becomes a function $1/{N^x}$. Here we consider three cases: $x=1,2,3$. The iterations are continued until $f_{n_c}(\lambda_0) \ge 0.99$ (or equivalently, $\epsilon \le 0.01$). In Fig.~\ref{grp:rand_anal}(a), we present $n_c$ versus $N$ graphs on a log-log scale. Each data is averaged over $1000$ simulations. The data are fitted to $\log{n_c} = A \log{N} + B$, and we find that ($A \simeq 1.130$, $B \simeq 0.103$) when $D=1/N$, ($A \simeq 2.017$, $B \simeq 0.330$) when $D=1/N^2$, and ($A \simeq 3.002$, $B \simeq 0.358$) when $D=1/N^3$. Note here that the fitting parameters $A$ are very well matched to the parameter $x$. These results allow us to estimate the value of $\alpha$: i.e. $\alpha \simeq 1$.

We also perform numerical simulations for the random Hamiltonian model to investigate $n_c$ for different accuracy (i.e. $1-\epsilon$) condition. In the simulations, we set as $D=1/N$, and consider three cases by taking $N=10$, $50$, and $100$. In Fig.~\ref{grp:rand_anal}(b), we give the graphs of $n_c$ versus $\epsilon$ on a log-log scale. Each data of point is also averaged over $1000$ simulations. The data are well fitted to $\log{n_c} = A\log{\epsilon} + B$. The fitting parameters $A$ and $B$ that we found are: ($A \simeq -0.186$, $B \simeq 0.811$) when $N=10$, ($A \simeq -0.192$, $B \simeq 1.605$) when $N=50$, and ($A \simeq -0.191$, $B \simeq 1.930$) when $N=100$. The parameters of $A$ are similar to $0.19$ in all the cases; thus, we estimate $\beta \simeq 0.19$.

%----------------------------
\section{Summary \& Discussion}\label{sec:5}

We have proposed a quantum algorithm to obtain the lowest energy eigenstate of a Hamiltonian. The process of the proposed algorithm could simply be regarded as a finite series of the transformations that consist of quantum Householder reflections and unitary of the system dynamics. Our algorithm is also assisted with an ancillary qubit system, similarly to the recently proposed method called ``Demon-like algorithmic (pseudo) cooling (DLAC)'', but, in our algorithm, the measurement is performed only once at the end of the algorithm to remove the ancillary qubit.  

In the theoretical analysis of Sec.~\ref{sec:3}, we proved that our algorithm can faithfully work for any given Hamiltonian. We also carried out numerical analyses in Sec.~\ref{sec:4}. First, we provided a numerical proof-of-principle demonstration for a simple Hamiltonian whose eigenvalues were equally spaced. The results showed a good agreement with our theoretical analysis, exhibiting the comparable behavior to the best `cooling' available with the DLAC method. We then considered a random Hamiltonian model for further analysis. We presumed that the total iterations $n_c$ required for an accuracy $f_{n_c}(\lambda_0) \ge 1-\epsilon$ would be proportional to ${\cal O}(c D^{-\alpha}\epsilon^{-\beta})$ with $\alpha,\beta \le 1$, where $D$ was difference between the two lowest eigenvalues, and $c$ is a constant factor. In the simulations, we estimated that $\alpha \simeq 1$ and $\beta \simeq 0.19 < 1$. 

In addition, we may compare our algorithm with the ``pseudo-cooling'' method \cite{Xu14}, which is also aimed to increase the fractions of the lower energy eigenstates, even though the temperature could not be well-defined in the process. In contrast to the method described in Ref.~\cite{Xu14} using a demon-like selective process by measurement, our algorithm does not contain any post-selection by measurement, but it requires a measurement at the end of the algorithm to remove the ancillary qubit system (See Appendix C for detailed argument).

Our algorithm may be useful to deal with many problems arising in the studies of complex systems, which may require to reach the lowest eigenstate of a given Hamiltonian. 

\begin{acknowledgements}
The authors thank Sunwhan Jo, Chanhyoup Lee, and Junghee Ryu for helpful discussions. We acknowledge the support of the Basic Science Research Program through National Research Foundation of Korea (NRF) funded by the Ministry of Science, ICT \& Future Planning (No. 2010-0018295 and No. 2010-0015059).
\end{acknowledgements}

%----------------------------
\appendix
%----------------------------

%----------------------------
\section{Proof of Eqs.~(\ref{eq:st_evolve})-(\ref{eq:factor_Wi})}

In order to prove Eqs.~(\ref{eq:st_evolve})-(\ref{eq:factor_Wi}), let us consider the transformation $\hat{T}_1$ at the first step. From Eq.~(\ref{eq:t_recur}), we can write $\hat{T}_1$ as
\begin{eqnarray}
\hat{T}_1 &=& \hat{R}_0 \hat{U} \hat{R}_0 \hat{U}^\dagger \nonumber \\
    &=& \big( \hat{\identity} - 2 \ket{\psi_0}\bra{\psi_0} \big)\big( \hat{\identity} - 2 \hat{U}\ket{\psi_0}\bra{\psi_0}\hat{U}^\dagger \big) \nonumber \\
    &=& \hat{\identity} - 2 \hat{U}\ket{\psi_0}\bra{\psi_0}\hat{U}^\dagger - 2 \ket{\psi_0}\bra{\psi_0} + 4 \bra{\psi_0}\hat{U}\ket{\psi_0}\ket{\psi_0}\bra{\psi_0}\hat{U}^\dagger.
\end{eqnarray}
Applying the above $\hat{T}_1$ to the composite initial state $\ket{\psi_0}$ [as in Eq.~(\ref{eq:init_st})], the output state $\ket{\psi_1}=\hat{T}_1 \ket{\psi_0}$ is computed such that
\begin{eqnarray}
\ket{\psi_1} &=& \ket{\psi_0} - 2 W_0^\ast \hat{U}\ket{\psi_0} - 2\ket{\psi_0} + 4\abs{W_0}^2 \ket{\psi_0} \nonumber \\
    &=& \left( 4\abs{W_0}^2 -1 \right)\ket{\psi_0} - 2W_0 \hat{U}\ket{\psi_0} \nonumber \\
    &=& \sum_{k=0}^{N-1} \frac{\gamma_{0,k}}{\sqrt{2}} \Big\{ \left[ \left(4\abs{W_0}^2 -1\right) - 2W_0^\ast e^{i \frac{\pi}{4}\tau\lambda_k} \right]\ket{0,\lambda_k} \nonumber \\
    && + \left[ \left(4\abs{W_0}^2 -1\right) - 2W_0^\ast e^{i \left(\frac{\pi}{2} - \frac{\pi}{4}\tau\lambda_k\right)}\right]\ket{1,\lambda_k} \Big\},
\label{eq:psi1_s}
\end{eqnarray}
where we let $W_0 = \bra{\psi_0}\hat{U}\ket{\psi_0}$. Here, we represent the complex number $W_0$ as the polar form,
\begin{eqnarray}
W_0 &=& \sum_{k=0}^{N-1} \abs{\gamma_{0,k}}^2 \left(\frac{ e^{i\frac{\pi}{4}\tau\lambda_k} + e^{i \left( \frac{\pi}{2} - \frac{\pi}{4}\tau\lambda_k\right)}}{2}\right) \nonumber \\
    &=& \sum_{k=0}^{N-1} \abs{\gamma_{0,k}}^2 \left( \frac{1}{\sqrt{2}} \cos{\frac{\pi \tau\lambda_k}{4}} + \frac{1}{\sqrt{2}} \sin{\frac{\pi \tau\lambda_k}{4}} \right) \left( \frac{1+ i}{\sqrt{2}} \right) \nonumber \\
    &=& \sum_{k=0}^{N-1} \abs{\gamma_{0,k}}^2 \left[ \cos{\frac{\pi\left( 1-\tau\lambda_k\right)}{4}} \right] e^{i \frac{\pi}{4}}.
\label{eq:w0}
\end{eqnarray}
Thus, using the above Eq.~(\ref{eq:w0}), we rewrite the result of Eq.~(\ref{eq:psi1_s}) as
\begin{eqnarray}
\ket{\psi_1} = \sum_{k=0}^{N-1} \frac{\gamma_{0,k}}{\sqrt{2}} \Big( \gamma_{1,k} \ket{0,\lambda_k} + \gamma_{1,k}^\ast \ket{1,\lambda_k} \Big),
\end{eqnarray}
where $\gamma_{1,k} = \left( 4\abs{W_0}^2 -1\right) - 2\abs{W_0} e^{-i \frac{\pi(1-\tau\lambda)}{4}}$.

Based on the results, we can generalize the above-described computations to the higher $i$th step ($i > 1$), using the Householder reflection $\hat{R}_{i-1} = \hat{\identity} - 2\ket{\psi_{i-1}}\bra{\psi_{i-1}}$ and unitary $\hat{U}$. In such generalization, we can easily find the expression of the $i$th output state $\ket{\psi_i}$ [as in Eq.~(\ref{eq:st_evolve})] and the coefficients $\gamma_{i,k}$ [as in Eq.~(\ref{eq:gamma})] with the factor $\abs{W_i}$ [as in Eq.~(\ref{eq:factor_Wi})]. 

%----------------------------
\section{The uniqueness of the solution Eq.~(\ref{eq:sol2})}

As mentioned in the main text, one may consider a more general situation, where
\begin{eqnarray}
\abs{\gamma_{i,k \le k'}}^2 \to 1,~\text{and},~ u_{k \le k'} \neq 0 ~\text{for}~i \to \infty.
\label{eq:a_sol}
\end{eqnarray}
Here, it is obvious that $\sum_{k \le k'}u_k = 1$. This can yield that $f_i(\lambda_0) \to u_0 < 1$ when $i \to \infty$, i.e. the probability $f_i(\lambda_0)$ of the lowest eigenstate cannot reach to $\simeq 1$. 

In such a general assumption, we prove that Eq.~(\ref{eq:sol2}) is the unique solution. To this end, we first give [using Eqs.~(\ref{eq:factor_Wi}) and (\ref{eq:assume_s})] 
\begin{eqnarray}
\abs{W_i} \to \sum_{k \le k'} u_k \cos\frac{\pi(1-\tau\lambda_k)}{4}~(\text{for}~i \to \infty).
\end{eqnarray}
Thus, for any specific eigenvalue $\lambda_{l}$ ($\le \lambda_{k'}$) and $i \to \infty$, we obtain [using Eq.~(\ref{eq:delta})]
\begin{eqnarray}
\Delta_{i,l} &\to& \sum_{k \le k'} u_k \cos\frac{\pi(1-\tau\lambda_k)}{4} - \cos\frac{\pi(1-\tau\lambda_l)}{4} \nonumber \\
    &=& \sum_{k \neq l \le k'} u_k \cos\frac{\pi(1-\tau\lambda_k)}{4} + (u_l - 1) \cos\frac{\pi(1-\tau\lambda_l)}{4} \nonumber \\
    &=& \sum_{k \neq l \le k'} u_k \left( \cos\frac{\pi(1-\tau\lambda_k)}{4} - \cos\frac{\pi(1-\tau\lambda_l)}{4} \right) \nonumber \\
    &=& \sum_{k \le k'} u_k \left( \cos\frac{\pi(1-\tau\lambda_k)}{4} - \cos\frac{\pi(1-\tau\lambda_l)}{4} \right),
\label{eq:delta_l}
\end{eqnarray}
where we used $\sum_{k \le k'}=1$, or $\sum_{k \neq l \le k'}u_k + (u_l -1) =0$. Here, we note that
\begin{eqnarray}
\abs{\gamma_{i,l}}^2 \to 1, ~\text{only when}~ \Delta_{i,l} \to 0 ~(\text{for}~i \to \infty),
\end{eqnarray}
which is verified by Eqs.~(\ref{eq:Q_ik}) and (\ref{eq:bound_wi}). Then, from Eq.~(\ref{eq:delta_l}), we find that the solution in Eq.~(\ref{eq:a_sol}) is possible when all the following conditions are satisfied:
\begin{eqnarray}
u_0 (c_0 - c_0) + u_1 (c_1 - c_0) + &\cdots& + u_{k'}(c_{k'} - c_0) = 0 ~(\text{for}~l=0), \nonumber \\
u_0 (c_0 - c_1) + u_1 (c_1 - c_1) + &\cdots& + u_{k'}(c_{k'} - c_1) = 0 ~(\text{for}~l=1), \nonumber \\
    &\vdots& \nonumber \\
u_0 (c_0 - c_{k'}) + u_1 (c_1 - c_{k'}) + &\cdots& + u_{k'}(c_{k'} - c_{k'}) = 0 ~(\text{for}~l=k'),
\label{eq:condi_a_sol}
\end{eqnarray}
where we let $c_k = \cos\frac{\pi(1-\tau\lambda_k)}{4}$ ($k = 0,1,\ldots,k'$) just for convenience. We rewrite the above conditions by adding all ($k'+1$) equations in Eq.~(\ref{eq:condi_a_sol}) as
\begin{eqnarray}
u_0 \left( (k'+1)c_0 - \sum_{k \le k'} c_k \right) + u_1 \left( (k'+1)c_1 - \sum_{k \le k'} c_k \right) + \cdots + u_{k'} \left( (k'+1)c_{k'} - \sum_{k \le k'} c_k \right) = 0.
\end{eqnarray}
Therefore, we can represent the condition for the existence of the solution Eq.~(\ref{eq:a_sol}) as
\begin{eqnarray}
\frac{1}{k'+1} \sum_{k \le k'} c_k = c_0 = c_1 = \cdots = c_{k'} ~\text{for}~ u_{k \le k'} \neq 0.
\label{eq:f_condi_a_sol}
\end{eqnarray}
However, we can directly see that this Eq.~(\ref{eq:f_condi_a_sol}) can {\em never} be satisfied, excepting the case $k' = 0$ and $u_0 = 1$ as in Eq.~(\ref{eq:sol2}) (as long as [{\bf P.1}] is true). \\

%----------------------------
\section{Comparison between our algorithm and ``pseudo-cooling'' method in Ref.~\cite{Xu14}} 

Here we discuss that our algorithm is quite distinct from the ``pseudo-cooling'' method proposed in Ref.~\cite{Xu14}, even though both methods may look similar. For example, let us consider the overall system Hamiltonian involving the ancillary qubit with the eigenvalues $E_k$ and their associated eigenstates $\ket{E_k}$, expressed as (for $k = 0,1, \ldots, N-1$)
\begin{eqnarray}
E_k = \frac{\pi}{4}\tau\lambda_k,~E_{2N-1-k} = \frac{\pi}{2}-\frac{\pi}{4}\tau\lambda_{k},
\label{eq:ek}
\end{eqnarray}
and
\begin{eqnarray}
\ket{E_k} = \ket{0, \lambda_k}, ~\ket{E_{2N-1-k}}=\ket{1,\lambda_k},\
\label{eq:ek_st}
\end{eqnarray}
where $E_0 < E_1 \le E_2 \le \ldots \le E_{2N-3} \le E_{2N-2} < E_{2N-1}$ [from Eq.~(\ref{eq:normalized_eigen})]. Using the set $\{E_k, \ket{E_k}\}$ ($k=0,1,\ldots,2N-1$) described above, we rewrite Eq.~(\ref{eq:uop}) as
\begin{eqnarray}
\hat{U} = \sum_{i=0}^{2N-1} e^{i \tau E_k} \ket{E_k}\bra{E_k},
\end{eqnarray}
with the lowest energy $E_0$ and the highest energy $E_{2N-1}$. Then we can rewrite Eq.~(\ref{eq:psi1_s}) as
\begin{eqnarray}
\ket{\psi_1} &=& \sum_{k=0}^{N-1}\frac{\gamma_{0,k}}{\sqrt{2}}  \underset{\text{($a$)}}{\left\{ \underbrace{\left[ \left(4\abs{W_0}^2 -1\right) - 2W_0^\ast e^{i \tau E_k} \right] \ket{E_k}}\right.} \nonumber \\
    && + \underset{\text{($b$)}}{\left.\underbrace{\left[ \left(4\abs{W_0}^2 -1\right) - 2W_0^\ast e^{i \tau E_{2N-1-k}} \right] \ket{E_{2N-1-k}}}\right\}},
\label{eq:psi1_2parts}
\end{eqnarray}
where we just replaced $\frac{\pi}{4}\tau\lambda_k$ and $\frac{\pi}{2}-\frac{\pi}{4}\tau\lambda_{k}$ in Eq.~(\ref{eq:psi1_s}) to $E_k$ [using Eq.~(\ref{eq:ek})], and did $\ket{0,\lambda_k}$ and $\ket{1,\lambda_k}$ to $\ket{E_k}$ [using Eq.~(\ref{eq:ek_st})]. Here, the terms `($a$)' and `($b$)' in Eq.~(\ref{eq:psi1_2parts}) describe the amplifications of the lower and the higher energy eigenstates respectively, so that they may look similar to the `cooled' and the `heated' parts, as in Ref.~\cite{Xu14}. However, we clarify that this is the case for the overall Hamiltonian (may not be of interest here). Furthermore, we do not use any measurement to `cool' the state of the (sub-)system Hamiltonian that we are really interested in. Therefore, our algorithm cannot be referred to as a `(pseudo) cooling', and thus the ancillary qubit system may play a different role in our algorithm. 

%\bibliography{common_fmmev.bib}   % name your BibTeX data base

\begin{thebibliography}{10}
\providecommand{\url}[1]{{#1}}
\providecommand{\urlprefix}{URL }
\expandafter\ifx\csname urlstyle\endcsname\relax
  \providecommand{\doi}[1]{DOI \discretionary{}{}{}#1}\else
  \providecommand{\doi}{DOI \discretionary{}{}{}\begingroup
  \urlstyle{rm}\Url}\fi

\bibitem{Liu07}
Y.K. Liu, M.~Christandl, F.~Verstraete, Phys. Rev. Lett. \textbf{98}, 110503
  (2007).

\bibitem{Fischer91}
K.H. Fischer, J.A. Hertz, \emph{Spin Glasses} (Cambridge University Press,
  Cambridge, 1991).

\bibitem{Rojas96}
R.~Rojas, \emph{Neural Networks: A Systematic Introduction} (Springer-Verlag,
  Berlin, 1996).

\bibitem{Albert02}
R.~Albert, A.L. Barab\'asi, Rev. Mod. Phys. \textbf{74}, 47 (2002).

\bibitem{Boykin02}
P.O. Boykin, T.~Mor, V.~Roychowdhury, F.~Vatan, R.~Vrijen, Proc. Natl Acad.
  Sci. USA \textbf{99}, 3388 (2002).

\bibitem{Fernandez04}
J.M. Fernandez, S.~Lloyd, T.~Mor, V.~Roychowdhury, Int. J. Quant. Comput.
  \textbf{2}, 461 (2004).

\bibitem{Verstraete09}
F.~Verstraete, M.M. Wolf, J.I. Cirac, Nature Phys. \textbf{5}, 633 (2009).

\bibitem{Xu14}
J.S. Xu, M.H. Yung, X.Y. Xu, S.~Boixo, Z.W. Zhou, C.F. Li, A.~Aspuru-Guzik,
  G.C. Guo, Nature Photonics \textbf{8}, 113 (2014).

\bibitem{Gershenfeld97}
N.A. Gershenfeld, I.L. Chuang, Science \textbf{275}, 350 (1997).

\bibitem{King98}
B.E. King, C.S. Wood, C.J. Myatt, Q.A. Turchette, D.~Leibfried, W.M. Itano,
  C.~Monroe, D.J. Wineland, Phys. Rev. Lett. \textbf{81}, 1525 (1998).

\bibitem{Roos99}
C.~Roos, T.~Zeiger, H.~Rohde, H.C. N\"agerl, J.~Eschner, D.~Leibfried,
  F.~Schmidt-Kaler, R.~Blatt, Phys. Rev. Lett. \textbf{83}, 4713 (1999).

\bibitem{James00}
D.F.V. James, arXiv:quant-ph/0003122  (2000).

\bibitem{Popp06}
M.~Popp, J.J. Garcia-Ripoll, K.G. Vollbrecht, J.I. Cirac, Phys. Rev. A
  \textbf{74}, 013622 (2006).

\bibitem{Christ07}
H.~Christ, J.I. Cirac, G.~Giedke, Phys. Rev. B \textbf{75}, 155324 (2007).

\bibitem{Lewenstein07}
M.~Lewenstein, A.~Sanpera, V.~Ahufinger, B.~Damski, A.~Sen(De), U.~Sen,
  Advances in Physics \textbf{56}, 243 (2007).

\bibitem{Bloch12}
I.~Bloch, J.~Dalibard, S.~Nascimb\'{e}ne, Nature Phys. \textbf{8}, 267 (2012).

\bibitem{Leff90}
H.S. Leff, A.F. Rex, \emph{MaxwellÕs Demon: Entropy, Information, Computing}
  (Princeton University Press, Princeton, NJ, 1990).

\bibitem{Lloyd97}
S.~Lloyd, Phys. Rev. A \textbf{56}, 3374 (1997).

\bibitem{Kosloff86}
R.~Kosloff, H.~Tal-ezer, Chem. Phys. Lett. \textbf{127}, 233 (1986).

\bibitem{Lehtovaara07}
L.~Lehtovaara, J.~Toivanen, J.~Eloranta, J. Comp. Phys. \textbf{221}, 148
  (2007).

\bibitem{Grover97}
L.K. Grover, Phys. Rev. Lett. \textbf{79}, 325 (1997).

\bibitem{Ivanov06}
P.A. Ivanov, E.S. Kyoseva, N.V. Vitanov, Phys. Rev. A \textbf{74}, 022323
  (2006).

\bibitem{Ivanov07}
P.A. Ivanov, B.T. Torosov, N.V. Vitanov, Phys. Rev. A \textbf{75}, 012323
  (2007).

\bibitem{Ivanov08}
P.A. Ivanov, N.V. Vitanov, Phys. Rev. A \textbf{77}, 012335 (2008).

\bibitem{Wannier60}
G.H. Wannier, Phys. Rev. \textbf{117}, 432 (1960).

\bibitem{Mendez88}
E.E. Mendez, F.~Agull\'o-Rueda, J.M. Hong, Phys. Rev. Lett. \textbf{60}, 2426
  (1988).

\bibitem{Voisin88}
P.~Voisin, J.~Bleuse, C.~Bouche, S.~Gaillard, C.~Alibert, A.~Regreny, Phys.
  Rev. Lett. \textbf{61}, 1639 (1988).

\bibitem{Abrams99}
D.S. Abrams, S.~Lloyd, Phys. Rev. Lett. \textbf{83}, 5162 (1999).

\bibitem{Yung12}
M.H. Yung, J.D. Whitefield, S.~Boixo, D.G. Tempel, A.~Aspuru-Guzik,
  arXiv:1203.1331  (2012).

\bibitem{Hioe81}
F.T. Hioe, J.H. Eberly, Phys. Rev. Lett. \textbf{47}, 838 (1981).

\end{thebibliography}

\end{document}